\documentclass[pra,aps,twocolumn]{revtex4-1}
\usepackage{bm,dcolumn,amsmath,graphicx,amsfonts,amssymb,lipsum,mathtools}
\usepackage{epsfig}
\usepackage{tikz}

\begin{document}

\title{ Using optical clock transitions in Cu II and Yb III for time-keeping and search for new physics}

\author{Saleh O. Allehabi$^1$}
\author{V. A. Dzuba$^1$}
\author{V. V. Flambaum$^{1,2}$}
\affiliation{$^1$School of Physics, University of New South Wales, Sydney 2052, Australia}
	
\affiliation{$^2$Helmholtz Institute Mainz, Johannes Gutenberg University, 55099 Mainz, Germany}
	
\date{\today}
	
\begin{abstract}
We study the $^1$S$_0 - ^3$D$_2$ and $^1$S$_0 - ^3$D$_3$ transitions in Cu~II and the $^1$S$_0 - ^3$P$^{\rm o}_2$ transition in Yb~III as 
possible candidates for the optical clock transitions.  A recently developed version of the configuration (CI) method, designed for a large number of electrons above closed-shell core, is used to carry out the calculation.
We calculate excitation energies, transition rates, lifetimes, scalar static polarizabilities of the ground and clock states, and blackbody radiation shift. 
We demonstrate that the considered transitions have all features of the clock transition leading to prospects of highly accurate measurements.
Search for new physics, such as time variation of the fine structure constant, is also investigated.
\end{abstract}
	
\maketitle

\section{Introduction}

Extremely high accuracy of the frequency measurements for the optical clock transitions naturally lead to the use of the transitions not only for time keeping but also for the search of the manifestations of new physics beyond the standard model, such as local Lorentz invariance (LLI) violation and time variation of the fine structure constant ($\alpha=e^2/\hbar c$)  (see, e.g. \cite{Ref1,Ref2,Ref3,Ref4,Ref5,Ref6,Ref7,Ref8}).
Oscillating variation of the fine structure constant  may be produced by  interaction of low mass scalar dark matter field with photon field (see, e.g., Refs. \cite{TilburgBudker,StadnikPRL,StadnikPRA,HeesGuena,Leefer}).  Therefore, the measurement of such  variation provides an efficient method to search for dark matter using atomic clocks, which have already provided improvement  of the constraints on the scalar - photon  interaction constants up to 15 orders of magnitude    \cite{TilburgBudker,StadnikPRL,StadnikPRA,HeesGuena,Leefer}.

The relative uncertainty of the frequency measurements for the best optical clocks is on the level of $10^{-18}$. For example, it is  9.4 $\times$10$^{-19}$ for Al$^+$~\cite{Ref4}, 3.0$\times$10$^{-18}$ for Yb$^+$~\cite{Ref6}, and 1$\times$10$^{-18}$ for Yb~\cite{Ybclock}. Unfortunately, most of working optical clocks are not very sensitive to new physics. Among the examples listed above, only Yb$^+$ clock transition is highly sensitive to variation of the fine structure  constant~\cite{CJP,Godun,Ref2} and to the LLI violation~\cite{SDzuba,Ref3}. Therefore, there is an ongoing search for new clock transitions which may combine high accuracy of the measurements with high sensitivity to new physics, e.g., to the time variation of the fine structure constant. 
One way of achieving this is to use highly charged ions~\cite{HCI1}. This is now a large area of research with very promising perspectives (see, e.g.~\cite{HCI2,HCI3,HCI4}).

Neutral or nearly neutral atoms are also considered. The important advantage of using them is that they are very well studied.
In some cases, new promising transitions can be found in atoms that are already used for a high accuracy atomic clock.
E.g., new transitions in Yb were recently suggested~\cite{DFS-Yb,SP-Yb} in addition to the currently used the  $^1$S$_0 - ^3$P$^{\rm o}_0$ clock transition. Clock transitions between metastable states in Yb~II have been  suggested in~\cite{Porsev}.  A good guide for finding atomic clock transitions sensitive to variations of $\alpha$ is to look for metastable states which are connected to the ground state via transitions that can be approximately considered as $s-d$, $s-f$ or $p-f$ single-electron transitions~\cite{Hg1}.
The $s-d$ transitions of this kind were considered in Cu, Ag and Au atoms in Ref.~\cite{CuAgAu}. 

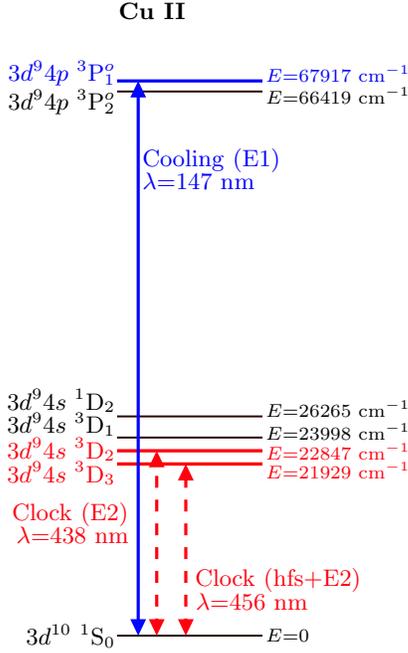
\begin{figure}[tb] 
		
\begin{center}
			
\tikzset{every picture/.style={line width=1.2pt}} 
			
\begin{tikzpicture}[x=1.pt,y=1.pt,yscale=-1,xscale=1]
			
			\draw (00,00) node [anchor=north west][inner sep=0.0pt]   [align=left] {\textbf{Cu~II}};
			
			\draw [color={rgb, 255:red, 0; green, 0; blue, 251 }  ,draw opacity=1 ][line width=1.2]    (0,30) -- (55,30) ; 
			
			\draw [color={rgb, 255:red, 41; green, 8; blue, 0 }  ,draw opacity=1 ][line width=0.75]    (0,34) -- (55,34) ;
			
			\draw [color={rgb, 255:red, 41; green, 8; blue, 0 }  ,draw opacity=1 ][line width=0.75]    (0,157) -- (55,157) ;
			
			\draw [color={rgb, 255:red, 41; green, 8; blue, 0 }  ,draw opacity=1 ][line width=0.75]    (0,165) -- (55,165) ;
			
			\draw [color={rgb, 500:red, 500; green, 2; blue, 27; brown, 0 }  ,draw opacity=1 ][line width=1.25]   (0,170) -- (55,170) ;
			
			\draw [color={rgb, 500:red, 500; green, 0; blue, 0; purple, 0 }  ,draw opacity=1 ][line width=1.25]   (0,175) -- (55,175) ;
			
			\draw [color={rgb, 255:red, 41; green, 8; blue, 0 }  ,draw opacity=1 ][line width=0.75]    (0,240) -- (55,240) ;
			
			\draw [color={rgb, 255:red, 4; green, 5; blue, 253 }  ,draw opacity=1 ]   (8,238) -- (8,32) ; 
			
			\draw [shift={(8,240)}, rotate = 270] [fill={rgb, 255:red, 4; green, 5; blue, 253 }  ,fill opacity=1 ][line width=0.08]  [draw opacity=0] (6.25,-3) -- (0,0) -- (6.25,3) -- cycle    ; 
			\draw [shift={(8,30)}, rotate = 90] [fill={rgb, 255:red, 4; green, 5; blue, 253 }  ,fill opacity=1 ][line width=0.08]  [draw opacity=0] (6.25,-3) -- (0,0) -- (6.25,3) -- cycle    ;

			\draw [color={rgb, 500:red, 500; green, 2; blue, 27; brown, 0}  ,draw opacity=1 ] [dash pattern={on 4.5pt off 4.5pt}]  (15,238) -- (15,172) ; 
			\draw [shift={(15,240)}, rotate = 270] [fill={rgb, 500:red, 500; green, 2; blue, 27; brown, 0 }  ,fill opacity=1 ][line width=0.08]  [draw opacity=0] (6.25,-3) -- (0,0) -- (6.25,3) -- cycle    ; 
			\draw [shift={(15,170)}, rotate = 90] [fill={rgb, 500:red, 500; green, 2; blue, 27; brown, 0 }  ,fill opacity=1 ][line width=0.08]  [draw opacity=0] (6.25,-3) -- (0,0) -- (6.25,3) -- cycle    ; 

			\draw [color={rgb, 500:red, 500; green, 2; blue, 27; brown, 0 }  ,draw opacity=1 ] [dash pattern={on 4.5pt off 4.5pt}]  (26,238) -- (26,177) ; 
			\draw [shift={(26,240)}, rotate = 270] [fill={rgb, 500:red, 500; green, 2; blue, 27; brown, 0 }  ,fill opacity=1 ][line width=0.01]  [draw opacity=0] (6.25,-3) -- (0,0) -- (6.25,3) -- cycle    ; 
			\draw [shift={(26,175)}, rotate = 90] [fill={rgb, 500:red, 500; green, 2; blue, 27; brown, 0 }  ,fill opacity=1 ][line width=0.01]  [draw opacity=0] (6.25,-3) -- (0,0) -- (6.25,3) -- cycle    ; 

			\draw (0,27) node [left][inner sep=0.0pt]  [color={rgb, 255:red, 0; green, 0; blue, 251 }  ,opacity=1 ]  {$3d^{9}4p$~$^3$P$^o_1$};
			
			\draw (0,38) node [left][inner sep=0.0pt]  [color={rgb, 255:red, 0; green, 0; blue, 0; black, 255 }  ,opacity=1 ]  {$3d^{9}4p$~$^3$P$^o_2$};
			
			\draw (0,151) node [left][inner sep=0.0pt]  [color={rgb, 255:red, 0; green, 0; blue, 0; black, 255 }  ,opacity=1 ]  {$3d^{9}4s$~$^1$D$_2$};
			
			\draw (0,160) node [left][inner sep=0.0pt]  [color={rgb, 255:red, 0; green, 0; blue, 0; black, 255 }  ,opacity=1 ]  {$3d^{9}4s$~$^3$D$_1$};
			
			\draw (0,169.5) node [left][inner sep=0.0pt]  [color={rgb, 500:red, 500; green, 2; blue, 27; brown, 0 }  ,opacity=1 ]  {$3d^{9}4s$~$^3$D$_2$};
			
			\draw (0,178.5) node [left][inner sep=0.0pt]  [color={rgb, 500:red, 500; green, 2; blue, 27; brown, 0 }  ,opacity=1 ]  {$3d^{9}4s$~$^3$D$_3$};
			
			\draw (0,240) node [left][inner sep=0.0pt]  [color={rgb, 255:red, 0; green, 0; blue, 0; black,255  }  ,opacity=1 ]  {$3d^{10}$~$^1$S$_0$};

			\draw (55,27) node [right][inner sep=0.75pt]  [font=\scriptsize] [color={rgb, 255:red, 0; green, 2; blue, 251 }  ,opacity=1 ]  {$E$=67917 cm$^{-1}$};
			
			\draw (55,35.5) node [right][inner sep=0.75pt]  [font=\scriptsize] [color={rgb, 255:red, 0; green, 2; blue, 0; black, 255 }  ,opacity=1 ]  {$E$=66419 cm$^{-1}$};
			
			\draw (55,154.) node [right][inner sep=0.75pt]  [font=\scriptsize] [color={rgb, 255:red, 0; green, 2; blue, 0; black, 255 }  ,opacity=1 ]  {$E$=26265 cm$^{-1}$};
			
			\draw (55,162.5) node [right][inner sep=0.75pt]  [font=\scriptsize] [color={rgb, 255:red, 0; green, 2; blue, 0; black, 255 }  ,opacity=1 ]  {$E$=23998 cm$^{-1}$};
			
			\draw (55,170.) node [right][inner sep=0.75pt]  [font=\scriptsize] [color={rgb, 500:red, 500; green, 2; blue, 27; brown, 0 }  ,opacity=1 ]  {$E$=22847 cm$^{-1}$};
			
			\draw (55,177.5) node [right][inner sep=0.75pt]  [font=\scriptsize] [color={rgb, 500:red, 500; green, 2; blue, 27; brown, 0 }  ,opacity=1 ]  {$E$=21929 cm$^{-1}$};
			
			\draw (55,240) node [right][inner sep=0.75pt]  [font=\scriptsize] [color={rgb, 255:red, 0; green, 0; blue, 0; black, 255 }  ,opacity=1 ]  {$E$=0};

			\draw (5,194) node [left][inner sep=0.0pt]  [color={rgb, 500:red, 500; green, 2; blue, 27; brown, 0 }  ,opacity=1 ]  {Clock (E2)};
			\draw (5,202) node [left][inner sep=0.0pt]  [color={rgb, 500:red, 500; green, 2; blue, 27; brown, 0}  ,opacity=1 ]  {$\lambda$=438 nm};
			
			\draw (61,219) node [anchor=center][inner sep=0.0pt]  [color={rgb, 500:red, 500; green, 2; blue, 27; brown, 0 }  ,opacity=1 ]  {Clock (hfs+E2)};
			\draw (51,227) node [anchor=center][inner sep=0.0pt]  [color={rgb, 500:red, 500; green, 2; blue, 27; brown, 0  }  ,opacity=1 ]  {$\lambda$=456 nm};
			
			\draw (9.,55) node [anchor=north west][inner sep=0.05pt]  [color={rgb, 255:red, 0; green, 0; blue, 255 }  ,opacity=1 ]  {Cooling (E1)  };
			
			\draw (9,64) node [anchor=north west][inner sep=0.05pt]  [color={rgb, 255:red, 0; green, 0; blue, 255 }  ,opacity=1 ]  {$\lambda$=147 nm  };
			

			\end{tikzpicture}
		\end{center}
		
		\caption{The energy diagram for the states of the Cu~II ion relevant for the optical ion clock. The electric dipole cooling transition is shown as a solid blue line, and the clock transitions are shown as short-dashed red lines.}
		\label{f:cu}
		
\end{figure}

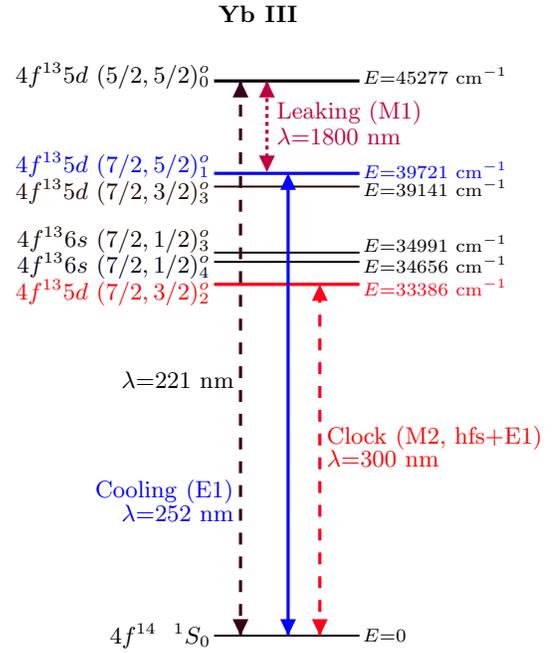
\begin{figure}[tb]
		
\begin{center}
			
			\tikzset{every picture/.style={line width=1.2pt}} 
			
			\begin{tikzpicture}[x=1.pt,y=1.pt,yscale=-1,xscale=1]

			\draw (00,00) node [anchor=north west][inner sep=1.pt]   [align=left] {\textbf{Yb~III  }};

			\draw [color={rgb, 500:red, 00; orange, 00; blue, ; pink, 00}  ,draw opacity=1 ][line width=1.2]    (0,30) -- (55,30) ; 

			\draw [color={rgb, 255:red, 0; green, 0; blue, 251 }  ,draw opacity=1 ][line width=1.2]    (0,65) -- (55,65) ; 
			
			\draw [color={rgb, 255:red, 41; green, 8; blue, 0 }  ,draw opacity=1 ][line width=0.75]    (0,70) -- (55,70) ;
			
			\draw [color={rgb, 255:red, 14; green, 6; blue, 0 }  ,draw opacity=1 ][line width=0.75]    (0,95) -- (55,95) ; 
			
			\draw [color={rgb, 255:red, 0; green, 2; blue, 0 }  ,draw opacity=1 ][line width=.75]    (0,98.5) -- (55,98.5) ;
			
			\draw [color={rgb, 255:red, 255; green, 2; blue, 27 }  ,draw opacity=1 ][line width=1.25]    (0,107) -- (55,107) ; 
			
			\draw [color={rgb, 255:red, 0; green, 0; blue, 0 }  ,draw opacity=1 ][line width=0.75]    (0,240) -- (55,240) ; 

			\draw [color={rgb, 255:red, 55; green, 2; blue, 27 }  ,draw opacity=1 ] [dash pattern={on 4.5pt off 4.5pt}]  (10,30) -- (10,240) ; 
			\draw [shift={(10,240)}, rotate = 270] [fill={rgb, 255:red, 55; green, 2; blue, 27 }  ,fill opacity=1 ][line width=0.08]  [draw opacity=0] (6.25,-3) -- (0,0) -- (6.25,3) -- cycle    ; 
			\draw [shift={(10,30)}, rotate = 90] [fill={rgb, 255:red, 55; green, 2; blue, 27 }  ,fill opacity=1 ][line width=0.08]  [draw opacity=0] (6.25,-3) -- (0,0) -- (6.25,3) -- cycle    ; 

			\draw [color={rgb, 255:red, 4; green, 5; blue, 253 }  ,draw opacity=1 ]   (28,240) -- (28,66) ; 
			\draw [shift={(28,240)}, rotate = 270] [fill={rgb, 255:red, 4; green, 5; blue, 253 }  ,fill opacity=1 ][line width=0.08]  [draw opacity=0] (6.25,-3) -- (0,0) -- (6.25,3) -- cycle    ; 
			\draw [shift={(28,65)}, rotate = 90] [fill={rgb, 255:red, 4; green, 5; blue, 253 }  ,fill opacity=1 ][line width=0.08]  [draw opacity=0] (6.25,-3) -- (0,0) -- (6.25,3) -- cycle    ;

			\draw [color={rgb, 500:purple, 500; red, 0 }  ,draw opacity=1 ] [dash pattern={on 1.5pt off 1.5pt}]  (20,62) -- (20,30) ; 
			\draw [shift={(20,64.5)}, rotate = 270] [fill={rgb, 500:purple,500; pink, 00; red, 0 }  ,fill opacity=1 ][line width=0.08]  [draw opacity=0] (6.25,-3) -- (0,0) -- (6.25,3) -- cycle    ; 
			\draw [shift={(20,30)}, rotate = 90] [fill={rgb, 500:purple,500; green, 0; red, 0 }  ,fill opacity=1 ][line width=0.08]  [draw opacity=0] (6.25,-3) -- (0,0) -- (6.25,3) -- cycle    ; 

			\draw [color={rgb, 255:red, 255; green, 2; blue, 27 }  ,draw opacity=1 ] [dash pattern={on 4.5pt off 4.5pt}]  (40,240) -- (40,109) ; 
			\draw [shift={(40,240)}, rotate = 270] [fill={rgb, 255:red, 255; green, 2; blue, 27 }  ,fill opacity=1 ][line width=0.08]  [draw opacity=0] (6.25,-3) -- (0,0) -- (6.25,3) -- cycle    ; 
			\draw [shift={(40,107)}, rotate = 90] [fill={rgb, 255:red, 255; green, 2; blue, 27 }  ,fill opacity=1 ][line width=0.08]  [draw opacity=0] (6.25,-3) -- (0,0) -- (6.25,3) -- cycle    ; 

			\draw (0,28) node [left][inner sep=0.75pt]  [color={rgb, 500:red, 00; green, 0; pink, 0 }  ,opacity=1 ]  {$4f^{13}5d$~$(5/2,5/2)^o_0$};
			
			\draw (23,37) node [anchor=north west][inner sep=0.05pt]  [color={rgb, 500:purple, 500; yellow, 0; blue, 0 }  ,opacity=1 ]  {Leaking (M1) };
			
			\draw (23,47) node [anchor=north west][inner sep=0.05pt]  [color={rgb, 500:purple, 500; green, 0; pink, 0 }   ,opacity=1 ]  {$\lambda$=1800 nm  };
			
			
			\draw (8,143) node [left][inner sep=0.0pt]  [color={rgb, 500:red, 00; green, 0; pink, 0 }   ,opacity=1 ]  {$\lambda$=221 nm  };

			\draw (0,62) node [left][inner sep=0.75pt]  [color={rgb, 255:red, 0; green, 0; blue, 251 }  ,opacity=1 ]  {$4f^{13}5d$~$(7/2,5/2)^o_1$};
			
			\draw (8,185.5) node [left][inner sep=0.05pt]  [color={rgb, 255:red, 0; green, 0; blue, 251 }  ,opacity=1 ]  {Cooling (E1) };
			
			\draw (8,193.5) node [left][inner sep=0.05pt]  [color={rgb, 255:red, 0; green, 0; blue, 251 }  ,opacity=1 ]  {$\lambda$=252 nm  };
			
			\draw (0,72) node [left][inner sep=0.75pt]  [color={rgb, 255:red, 35; green, 6; blue, 0 }  ,opacity=1 ]  {$4f^{13}5d$~$(7/2,3/2)^o_3$};
			
			\draw (0,90) node [left][inner sep=0.75pt]  [color={rgb, 255:red, 19; green, 7; blue, 0 }  ,opacity=1 ]  {$4f^{13}6s$~$(7/2,1/2)^o_3$};
			
			\draw (0,100) node [left][inner sep=.75pt]  [color={rgb, 255:red, 0; green, 2; blue, 27 }  ,opacity=1 ]  {$4f^{13}6s$~$(7/2,1/2)^o_4$};
			
			\draw (0,110) node [left][inner sep=.75pt]  [color={rgb, 255:red, 255; green, 0; blue, 0 }  ,opacity=1 ]  {$4f^{13}5d$~$(7/2,3/2)^o_2$};
			
			\draw (42,160) node [anchor=north west][inner sep=0.05pt]  [color={rgb, 255:red, 255; green, 0; blue, 0 }  ,opacity=1 ]  {Clock (M2, hfs+E1)  };
			
			\draw (42,169) node [anchor=north west][inner sep=0.05pt]  [color={rgb, 255:red, 255; green, 0; blue, 0 }  ,opacity=1 ]  {$\lambda$=300 nm };
			
			\draw (0,240) node [left][inner sep=.75pt]  [color={rgb, 255:red, 0; green, 0; blue, 0 }  ,opacity=1 ]  {$4f^{14}$~ $^{1} S_{0}$};

			\draw (55,28) node [right][inner sep=0.75pt]  [font=\scriptsize] [color={rgb, 500:red, 00; green, 2; pink, 0 }  ,opacity=1 ]  {$E$=45277 cm$^{-1}$};
			
			\draw (55,63.5) node [right][inner sep=0.75pt]  [font=\scriptsize] [color={rgb, 255:red, 0; green, 2; blue, 251 }  ,opacity=1 ]  {$E$=39721 cm$^{-1}$};
			
			\draw (55,70.5) node [right][inner sep=0.75pt]  [font=\scriptsize]  {$E$=39141 cm$^{-1}$};
			
			\draw (55,91.5) node [right][inner sep=0.75pt]  [font=\scriptsize]  {$E$=34991 cm$^{-1}$};
			
			\draw (55,99.5) node [right][inner sep=0.75pt]  [font=\scriptsize]  {$E$=34656 cm$^{-1}$};
			
			\draw (55,108) node [right][inner sep=0.75pt]  [font=\scriptsize] [color={rgb, 255:red, 255; green, 2; blue, 27 }  ,opacity=1 ]  {$E$=33386 cm$^{-1}$};
			
			\draw (55,240) node [right][inner sep=.75pt]  [font=\scriptsize]  {$E$=0};

			\end{tikzpicture}
		\end{center}
		
	\caption{The energy diagram for the states of the Yb~III ion relevant for the optical ion clock. The electric dipole cooling transition is shown as a solid blue line, the clock transition is shown as a short-dashed red line, and the purple dotted lines show the leakage transition. }
		\label{f:yb}
		
	\end{figure}

In the present paper, we consider the $^1$S$_0 - ^3$D$_2$ and $^1$S$_0 - ^3$D$_3$ transitions in Cu~II and the $^1$S$_0 - ^3$P$^{\rm o}_2$ transition in Yb~III (see Figs. \ref{f:cu} and \ref{f:yb}). Transitions in Cu~II are the $s-d$ transitions, the transition in Yb~III is the $s-f$ transition. In our early work~\cite{CJP}, we suggested to use the $4f^{14} \ ^1$S$_0 - 4f^{13}5d \  (\frac{5}{2},\frac{5}{2})^{\rm o}_0$ in Yb~III for the search of the variation of the fine structure constant. The prospect for precision measurement of the frequency of this transition was considered in a recent paper~\cite{Naoki}. However, this transition has an important drawback. 
There is a decay channel via magnetic dipole transition (M1) into lower-lying state  $4f^{13}5d \  (\frac{7}{2},\frac{5}{2})^{\rm o}_1$. 
This may make the considered transition to be not sufficiently narrow  to ensure high accuracy of the measurements.
This problem was not discussed in \cite{CJP} or \cite{Naoki}. 
In the present paper, we consider a different transition, a transition from the ground state to the first excited state  $4f^{13}5d \  (\frac{7}{2},\frac{3}{2})^{\rm o}_2$. This is a very narrow transition with a similar sensitivity to the variation of the fine structure constant. We demonstrate that it has all features of the atomic clock transition.

Several studies have analyzed the energies and transition probabilities for both ions, Cu~II~\cite{Pinnington,Andersson,Dong} and Yb~III~\cite{Safronova,Zhang}  theoretically and experimentally (see also \cite{NIST} and references therein). This gives us an opportunity to compare results to have confidence in the accuracy of the analysis. None of the previous studies focused on transitions  in Cu~II and Yb~III in sufficient details to study their suitability for time-keeping and searching for  new physics. 	

\section{Method of calculation}
	
	As can be seen from the spectra of the Cu~II and Yb~III ions, the excited states of the Cu~II ion have an open $3d$ shell and the excited states of the Yb~III ion have an open $4f$ shell. Therefore, to perform the electron structure calculations for both ions, the recent version of the configuration interaction (CI) method was used, which has been designed to deal with a large number of valence electrons~\cite{CIPT}. The method combines CI with perturbation theory (PT) and is called the CIPT method. The method reduces the size of the effective CI matrix by neglecting the off-diagonal matrix elements between high-energy basis states and reducing their contribution to PT corrections to the matrix elements between low-energy basis states.

The eigenvalues $E$ and eigenstates~ $\psi$ can be found by solving the CI equations with the effective $H^{CI}$ matrix
	
	\begin{equation}
	\left(H^{CI}-EI\right)  \psi = 0,
	\label{e:CI} 
	\end{equation}
where $I$ is the unit matrix.
Matrix elements of the effective CI matrix contain PT-type corrections from the high-energy states
	  
\begin{equation}
	  \langle a|H^{CI}|b\rangle\rightarrow\langle a|H^{CI}|b\rangle+\sum_{h}\frac{\langle a|H^{CI}|h\rangle\langle h|H^{CI}|b\rangle}{E-E_{h}}.
	  \label{e:CIPT}
\end{equation}
Here \textit{a} and \textit{b} are  low-energy states, and $E_{h}$ is the diagonal
matrix element between high-energy states, ($E_{h}=\langle h|H^{\rm CI}|h \rangle \delta_{h}$).
To produce a set of complete single-electron basis states for both ions, we start the calculations with the Dirac-Hartree-Fock (DHF)  method in the V$^{N}$ approximation with all atomic electrons included.  It seems to be natural to start from the [Ar]$3d^{10}$ configuration for Cu~II and the  [Xe]$4f^{14}$ configuration for Yb~III. However, such a choice of initial approximation is good for calculating the ground states of the ions. Since we need to calculate excited states as well, which have excitations from the $3d$ or $4f$ subshell, the choice of initial approximation is not obvious, and it is dictated by the accuracy of the final results. It turns out that the best results are obtained if we start from  the [Ar]$3d^{9}4s$ configuration for Cu~II and the  [Xe]$4f^{14}$ configuration for Yb~III. 
	
The single-electron basis states are then constructed using B-splines~\cite{Johnson_Bspline,Johnson_Bspline2} with forty B-splines states of the order k=9 in a box of the radius R$_{max}$ = 40$\textit{a}_B$ with orbital angular momentum 0~$\leq$~\textit{l}~$\leq$~4.
	
To carry out the calculations of the transition amplitudes and hyperfine structure (hfs), we use the time-dependent Hartree-Fock (TDHF) method~\cite{CPM},
which is equivalent to the random-phase approximation (RPA).
The RPA equations can be written as
\begin{equation}\label{e:RPA}
	(\hat H^{RHF}-\epsilon_c)\delta\psi_c=-(\hat d+\delta V^{N})\psi_c.
\end{equation}
Here $\hat d$ refers to the operator of an external field, which can be any field, which is sufficiently weak to be considered in linear approximation. 
$\epsilon_c$ is the energy of electron state $c$, $\psi_c$ is the state wave function, and $\delta V^{N}$ denotes the correction to the self-consistent  potential caused by the effect of an external field. Equations (\ref{e:RPA}) are solved self-consistently for all states $c$ in the core.
Then matrix elements for valence states are calculated using the expression 
\begin{equation}\label{e:Ab}
	A_{b \rightarrow a}\equiv\langle \psi_a|\hat d+\delta V^{N}|\psi_b\rangle.
\end{equation}
The electric dipole (E1),  magnetic dipole (M1), electric quadrupole (E2), magnetic quadrupole (M2), and electric octupole (E3) transition probabilities (in atomic units) from upper state\textit{ b} to lower state \textit{a} can be written as 
	
	\begin{equation}\label{e:Td}
	T_{\rm E1,M1} = \frac{4}{3}(\alpha\omega)^3 \frac{A^2_{E1,M1}}{2J_b+1},
	\end{equation}
	
	\begin{equation}\label{e:Tq2}
	T_{\rm E2,M2} = \frac{1}{15}(\alpha\omega)^5 \frac{A^2_{E2,M2}}{2J_b+1},
	\end{equation}
	
		\begin{equation}\label{e:Tq}
	T_{\rm E3} = 0.001 69(\alpha\omega)^7 \frac{A^2_{E3}}{2J_b+1}.
	\end{equation}
Here $\alpha$ is the fine structure constant, $\omega$ is the energy difference between the lower and upper states, \textit{A} is the transition amplitude (\ref{e:Ab}), $J_b$ is the total angular momentum of the upper state \textit{b}. Note that magnetic amplitudes $A_{M1, M2}$ contain the Bohr magneton $\mu_B$ ($\mu_B = \alpha/2 \approx 3.65 \times 10^{-3}$ in atomic units). 
For some strongly forbidden transitions leading contribution comes from electromagnetic transitions mediated by the  hyperfine interaction. 
Clock transitions in $^{63,65}$Cu~II and $^{171,173}$Yb~III are good examples of such transitions. 
The transition amplitude  is
\begin{equation}\label{e:HF+E2}
	\begin{aligned}
	A_{\rm hfs-E1,E2}(b \rightarrow a) 
	=\sum_{n}\left(\frac{\langle a|{A_{\rm hfs}}| n\rangle \langle n|{A_{\rm E1,E2}}|b \rangle}{\Delta E} \right.\\
	+\left. \frac{\langle b| {A_{\rm hfs}}| n\rangle\langle n|{A_{\rm E1,E2}}| a\rangle}{\Delta E}\right).
	\end{aligned}
\end{equation}
Here $A_{\rm hfs}$ is the operator of the magnetic dipole or electric quadrupole hfs interaction, 
and ${A_{\rm E1,E2}}$ are the operators of the E1 and E2 transitions. Summation in (\ref{e:HF+E2}) goes over a complete set of intermediate states $|n\rangle$ (for more details, see, e.g. ~\cite{HFI,HFI2,Andersson}). In practice, it is usually sufficient to include few close states into the summation over $n$.
For example, the leading contribution to the transition amplitude of the 
$^1$S$_0$ - $^3$D$_3$ clock transition in Cu~II comes from the electric quadrupole transition mediated by the magnetic dipole hfs interaction.
It is sufficient to include three intermediate states into the summation, the $3d^94s$ $^3$D$_2$, $^1$D$_2$, and $^3$D$_1$ states.
Then Eq.~(\ref{e:HF+E2}) becomes
\begin{equation}\label{e:hfs-Cu}
	\begin{aligned}
	A_{\mathrm{hfs}-E2}({3d^{9}4s~\rm ^{3}D_{3}} \rightarrow {3d^{10}~\rm ^{1}S_{0}})= \qquad\\
	 \frac{\langle {\rm ^{3}D_{3}}|\hat{H}_{\mathrm{hfs}}| {\rm ^{3}D_{2}}\rangle\langle {\rm ^{3}D_{2}}|\hat{E2}| {\rm ^{1}S_{0}}\rangle }{\Delta E} +\\
	 \frac{\langle {\rm ^{3}D_{3}}|\hat{H}_{\mathrm{hfs}}| {\rm ^{1}D_{2}}\rangle\langle {\rm ^{1}D_{2}}|\hat{E2}| {\rm ^{1}S_{0}}\rangle}{\Delta E} +\\
	  \frac{\langle {\rm ^{3}D_{3}}|\hat{E2}| {\rm ^{3}D_{1}}\rangle\langle {\rm ^{3}D_{1}}|\hat{H}_{\mathrm{hfs}}| {\rm ^{1}S_{0}}\rangle}{\Delta E}
	\end{aligned}
	\end{equation}
For the 2-1 clock transition in the Yb~III ion, the hyperfine-induced E1 transition amplitude is expressed as
\begin{equation}
\begin{aligned}\label{e:hfs-Yb}
	A_{\mathrm{hfs}-E1}({4f^{13}5d~\rm ^{3}P^o_{2}} \rightarrow {4f^{14}~\rm ^{1}S_{0}})=\qquad\\
	\frac{\langle {\rm ^{3}P^o_{2}}|\hat{H}_{\mathrm{hfs}}| {\rm ^{3}P^o_{1}}\rangle\langle {\rm ^{3}P^o_{1}}|\hat E1| {\rm ^{1}S_{0}}\rangle}{\Delta E} 
	\end{aligned}
\end{equation}
To find corresponding transition rates, we use Eq.($\ref{e:Td}, \ref{e:Tq2}$), replacing ${A_\mathrm{E1}}$ by $A_{\mathrm{hfs}-\mathrm{E}1}$ in Eq.(\ref{e:Td}), and ${A_\mathrm{E2}}$ by $A_{\mathrm{hfs}-\mathrm{E}2}$ in Eq.(\ref{e:Tq2}).
	
Radiative lifetimes $ \tau_b$ of each excited state $b$ can be obtained as
\begin{equation}
\tau_b =  1/\sum_a T_{ab}
\end{equation}
where summation goes over all possible transitions to lower states $a$.

		\begin{table*}
		
		\caption{\label{t:Energy}
	Excitation energies ($E$, cm$^{-1}$) and lifetimes ($\tau$) for some low  states of Cu~II and Yb~III ions
	.} 
		\begin{ruledtabular}
			\begin{tabular}{cccccccll}
				&&&
				\multicolumn{3}{c}{Energy [cm$^{-1}$]}&
				\multicolumn{3}{c}{Lifetime}\\
				
				\cline{4-6}
				\cline{7-9}
				
				&&&&
				\multicolumn{2}{c}{Other}&&
				\multicolumn{2}{c}{Other}\\
				
				\cline{5-6}
				\cline{8-9}
				
				\multicolumn{1}{c}{N}& 
				\multicolumn{1}{c}{Conf.}&
				\multicolumn{1}{c}{Term}&
				
				\multicolumn{1}{c}{Present }&
				\multicolumn{1}{l}{NIST \cite{NIST}}&
				\multicolumn{1}{c}{Cal. }&

				\multicolumn{1}{c}{Present }&
				\multicolumn{1}{c}{Exp.}&
				\multicolumn{1}{c}{Cal.}\\

				\hline
				\multicolumn{9}{c}{\textbf{Cu~II}}\\
				
				1 & $3d^{10}$&$^1${S}$_{0}$&0&0&0\footnotemark[1]&$\infty$&&\\
				2 &$3d^{9}4s$&$^3${D}$_{3}$&21932&21929&22469\footnotemark[1]&2.5$\cdot$10$^{+8}$ s&&\\
				3&& $^3${D}$_{2}$ &22733&22847&23381\footnotemark[1]&7.8 s&&\\
				4&& $^3${D}$_{1}$ &23705&23998&24495\footnotemark[1]&&&\\
				5 &$3d^{9}4s$&$^1${D}$_{2}$&25833&26265&26840\footnotemark[1]&&&\\
				6 & $3d^{9}4p$& $^3${P}$^{\rm o}_{2}$ &66623&66419&66984\footnotemark[2]&&&\\
				7 && $^3${P}$^{\rm o}_{1}$ &67922&67917&68703\footnotemark[2]&2.2 ns&2.36$\pm$0.05 ns\footnotemark[3]&2.39 ns, 2.21 ns\footnotemark[2]\\
				\\

				\multicolumn{9}{c}{\textbf{Yb~III}}\\
				1 & $4f^{14}$&$^1${S}$_{0}$&0&0&0\footnotemark[4]&$\infty$&&\\
				2 & $4f^{13}5d$&(7/2,3/2)$^{\rm o}_{2}$$\equiv$$^3${P}$\rm ^o_{2}$&29208&33386& 39755\footnotemark[4]& 2000  s && 6017 s\footnotemark[4] \\
				3 & $4f^{13}5d$&(7/2,3/2)$^{\rm o}_{3}$$\equiv$$^3${D}$\rm ^o_{3}$&33839&39141& 44429\footnotemark[4]&&&\\
				4 & $4f^{13}6s$&(7/2,1/2)$^{\rm o}_{4}$$\equiv$$^3${F}$^{\rm o}_{4}$&35000&34656& 36336\footnotemark[4]&&&\\
				5 & $4f^{13}6s$&(7/2,1/2)$^{\rm o}_{3}$$\equiv$$^1${F}$^{\rm o}_{3}$&36418&34991& 36764\footnotemark[4]&&&\\
				6 & $4f^{13}5d$&(7/2,5/2)$^{\rm o}_{1}$$\equiv$$^3${P}$^{\rm o}_{1}$&35288&39721& 39762\footnotemark[4]& 250 ns &230(20) ns\footnotemark[5]& 166 ns, 270 ns\footnotemark[5]\\
				
				&&&&&&&& 181 ns\footnotemark[4]\\
				7 & $4f^{13}5d$&(5/2,5/2)$^{\rm o}_{0}$$\equiv$$^3${P}$^{\rm o}_{0}$&41059&45277& 49469\footnotemark[4]&  0.133 s & &0.1490 s\footnotemark[4]\\

			\end{tabular}
		
			\footnotetext[1] { Ref.~\cite{Andersson}.}
			\footnotetext[2] { Ref.~\cite{Dong}; the first calculated value was obtained using the length gauge and the second calculated value was obtained using the velocity gauge.}
			\footnotetext[3] { Ref.~\cite{Pinnington}.}
			\footnotetext[4] { Ref.~\cite{Safronova}; the value was obtained using the RMBPT method.}
			\footnotetext[5] { Ref.~\cite{Zhang}; the first calculated value was obtained using the RHF method+CP effects; the second calculated value was obtained using same procedure with including 5\textit{s}, 5\textit{p}, and 4\textit{f} to the CP effects.}


		\end{ruledtabular}
	\end{table*}

	\section{Results}

\subsection{Energy levels, transition probabilities, and lifetimes.}
	
	
Table~\ref{t:Energy} presents calculated energy levels and lifetimes of the low-energy states of Cu~II and Yb~III ions compared with experimental data and other calculations. The lifetimes were calculated using transition probabilities presented in Table~\ref{t:Tran}.
The results for the energies are in sufficiently good agreement with experimental data from NIST.  The average difference between the NIST and calculated data for Cu~II is  $\sim$ 100 cm$^{-1}$, while for Yb~III, the difference  is $\sim$ 4000 cm$^{-1}$.
Note that different sources present different state labeling for Yb~III (see, e.g., \cite{Naoki,NIST}). Therefore, for the sake of easy comparison, we present in the table state labeling based on both commonly used schemes, the $J-J$ and $L-S$ schemes. 
	



	\begin{table*}
	
	\caption{\label{t:Tran} Transition amplitudes (\textit{A}, a.u.) and transition probabilities (T, 1/s) evaluated with NIST frequencies for some low  states of Cu~II and Yb~III ions. Semi.$\equiv$ Semiempirical.} 
	
\begin{ruledtabular}
	\begin{tabular}{cclllllll}
		&&
		\multicolumn{2}{c}{($\omega$), NIST \cite{NIST} }&
		\multicolumn{2}{c}{Present}&
		\multicolumn{3}{c}{Other, T [s$^{-1}$]}\\
		\cline{3-4}
		\cline{5-6}
		\cline{7-9}
		
		\multicolumn{1}{c}{Transition}& 
		\multicolumn{1}{c}{Type}&
		\multicolumn{1}{c}{ [cm$^{-1}$]}&
		\multicolumn{1}{c}{ [a.u.]}&

		\multicolumn{1}{c}{\textit{A} [a.u]}&
			\multicolumn{1}{c}{T [s$^{-1}$]}&
			\multicolumn{1}{c}{Exp.~\cite{Pinnington}}&
			\multicolumn{1}{c}{Semi.~\cite{Dong}}&
			\multicolumn{1}{c}{Cal.}\\

			\hline
		\multicolumn{9}{c}{\textbf{Cu~II}}\\
		
		2$\leftrightarrow$1 & hfs-E2&21929&0.0999&-2.204$\cdot$10$^{-4}$\footnotemark[1]&3.935$\cdot$10$^{-9}$\footnotemark[1]&&&5.550$\cdot$10$^{-9}$\footnotemark[3] \\
		
		2$\leftrightarrow$1 & hfs-E2&21929&0.0999&-2.361$\cdot$10$^{-4}$\footnotemark[2]&4.516$\cdot$10$^{-9}$\footnotemark[2]&&&6.383$\cdot$10$^{-9}$\footnotemark[3] \\
		
		3$\leftrightarrow$1 & E2&22847&0.1041&0.890&11.03$\cdot$10$^{-2}$&&
		&10.4$\cdot$10$^{-2}$, 15.7$\cdot$10$^{-2}$\footnotemark[3]
		\\
		
		3$\leftrightarrow$2 &M1&918&0.0042&2.070$\mu_B$& 1.790$\cdot$10$^{-2}$&&&1.70$\cdot$10$^{-2}$\footnotemark[4]\\
		
		5$\leftrightarrow$1 & E2&26265&0.1197&-2.727&2.080& && 1.937, 2.687\footnotemark[3]\\
		
		7$\leftrightarrow$1&E1& 67917&0.3095&-0.182&6.689$\cdot$10$^{+6}$&11.3$\cdot$10$^{+6}$& 8.5$\cdot$10$^{+6}$&   7.6$\cdot$10$^{+6}$, 7.7$\cdot$10$^{+6}$\footnotemark[5]\\
		
		7$\leftrightarrow$2 & E3&45988&0.2095&-0.346&5.459$\cdot$10$^{-8}$&&&\\ 
		
		7$\leftrightarrow$3 & E1&45069&0.2054&2.489&3.826$\cdot$10$^{+8}$&3.419$\cdot$10$^{+8}$ &3.474$\cdot$10$^{+8}$ &3.425 $\cdot$10$^{+8}$, 3.628$\cdot$10$^{+8}$\footnotemark[5]\\
		
		7$\leftrightarrow$4 &E1&43918&0.2001&1.097&6.875$\cdot$10$^{+7}$&6.29$\cdot$10$^{+7}$ &6.35$\cdot$10$^{+7}$&   6.29 $\cdot$10$^{+7}$, 7.16$\cdot$10$^{+7}$\footnotemark[5]\\
		
		7$\leftrightarrow$5&E1& 41652&0.1898&0.379&7.014$\cdot$10$^{+6}$&7.7$\cdot$10$^{+6}$ & 8.5$\cdot$10$^{+6}$ & 6.8$\cdot$10$^{+6}$, 7.6$\cdot$10$^{+6}$\footnotemark[5]\\

		7$\leftrightarrow$6 & E2&1498&0.0068&0.656&1.211$\cdot$10$^{-7}$&&&\\
		
		\\
		\multicolumn{9}{c}{\textbf{Yb~III}}\\
		2$\leftrightarrow$1 & M2&33386&0.1521&5.612$\mu_B$&3.895$\cdot$10$^{-4}$&&& 1.662$\cdot$10$^{-4}$\footnotemark[6]\\
		
		2$\leftrightarrow$1 &hfs-E1&33386&0.1521& 2.581$\cdot$10$^{-6}$\footnotemark[7]&1.004$\cdot$10$^{-4}$\footnotemark[7]&&&\\
		
		2$\leftrightarrow$1 & hfs-E1&33386&0.1521& -6.777$\cdot$10$^{-7}$\footnotemark[8]&6.918$\cdot$10$^{-6}$\footnotemark[8]&&&\\
		
		6$\leftrightarrow$1 & E1&39721&0.1810&0.308&4.015$\cdot$10$^{+6}$&&& 5.524$\cdot$10$^{+6}$\footnotemark[6]\\
		
		6$\leftrightarrow$2 & M1&6335&0.0289&1.583$\mu_B$&5.726&&& 5.702\footnotemark[6]\\
		
		6$\leftrightarrow$3 & E2&580&0.0026&0.909& 2.017$\cdot$10$^{-9}$&&&\\
		
		
		6$\leftrightarrow$5& E2 &4730&0.0216 &7.202&4.579$\cdot$10$^{-3}$& && 3.516$\cdot$10$^{-2}$\footnotemark[6]\\

		7$\leftrightarrow$2 & E2&11891	&0.0542&0.529&7.442$\cdot$10$^{-3}$&&&5.209$\cdot$10$^{-3}$\footnotemark[6]\\
		7$\leftrightarrow$6 & M1&5556&0.0253&1.275$\mu_B$&7.523&&& 6.706\footnotemark[6]\\
		\end{tabular}
\footnotetext[1]{$^{63}$Cu.}		
\footnotetext[2]{$^{65}$Cu.}		
		\footnotetext[3] { Ref.~\cite{Andersson}; for the 3-1 transition, the first value was obtained using the Babushkin gauge and the second value was obtained using the Coulomb gauge.}

\footnotetext[4] { Ref.~\cite{Garstang}.}
\footnotetext[5] { Ref.~\cite{Dong}; the first calculated value was obtained using the length gauge and the second calculated value was obtained using the velocity gauge.}

\footnotetext[6] { Ref.~\cite{Safronova}; the value was obtained using the RMBPT method.}
\footnotetext[7]{$^{171}$Yb.}		
\footnotetext[8]{$^{173}$Yb.}		
		
	\end{ruledtabular}
\end{table*}

Table~\ref{t:Tran} presents calculated transition amplitudes and transition rates and compares them to the experimental data and other theoretical values.
Lifetimes of the states calculated using transition rates from Table~\ref{t:Tran} are presented in Table~\ref{t:Energy}.
As can be seen from the tables, the present results for the Cu~II ion are in good agreement with the experimental data and other calculations.
For the transition between the first excited state $3d^94s$ $^3$D$_3$ and the ground state, the dominating contribution comes from the 
hfs-induced electric quadrupole transition (see Eq.~(\ref{e:hfs-Cu})).
This transition was studied before in Ref.~\cite{Andersson} using the same strategy. However, the authors calculated the transition rates separately for different hfs components of the states. Their results for transition rates for $^{63,65}$Cu~II for different values of the total angular momentum $F$ ($\mathbf{F}=\mathbf{J}+\mathbf{I}$, where $I$ is unclear spin) range
between (3.10$\cdot$10$^{-12}$ $-$ 4.29$\cdot$10$^{-9}$ s$^{-1}$) and (3.19$\cdot$10$^{-12}$ $-$ 1.06$\cdot$10$^{-8}$ s$^{-1}$), respectively, indicating good agreement with our calculations.
	
For the transition rates of Yb~III ion, we compared our results with the theoretical values of Safronova \textit{et al.}~\cite{Safronova}. 
 They carried out theoretical calculations using 
the second-order relativistic many-body perturbation theory (RMBPT). The results are in reasonably good agreement with our calculations. 
The most noticeable disagreement is about two times difference in the $M2$ transition rate between the clock and ground states. 
Given that hfs-induced $E1$ transition also gives a significant contribution into the transition rate, and this contribution was not considered in Ref.~\cite{Safronova}, the total difference in the lifetime of the clock state is about three times (see Table~\ref{t:Energy}).
	
The data on lifetimes for the states of both ions are presented in Table~\ref{t:Energy} . The present results are compared with experimental and other theoretical calculations. For the Yb~III ion,  Zhang \textit{et al.}~\cite{Zhang}  have obtained the lifetime result for the $4f^{13}5d$ $^3${P}${\rm ^o_1}$ state both experimentally and theoretically. They performed the calculations using two variations of the relativistic Hartree-Fock (RHF) method of Cowan~\cite{Cowan}, which differ by the ways of inclusion of the core polarisation (CP) effect.


\subsection{Polarizabilities and Blackbody Radiation Shifts}
		
Static scalar polarizability $\alpha_v(0)$ of an atom in state $v$ is given by
\begin{equation}\label{e:pol}
	\alpha_v(0)=\dfrac{2}{3(2J_v+1)}\sum_{n}\frac{A_{vn}^2}{\omega_{vn}},
\end{equation}
where $J_v$ is the total angular momentum of  state $v$,  $A_{vn}$ are the amplitudes (reduced matrix elements) of the electric dipole transitions, $\omega_{vn}$ is the frequency of the transition. Eq.~(\ref{e:pol}) is valid when all wave functions $v$ and $n$ are many-electron wave functions of the whole atom.
It can also be used to calculate valence contributions to the polarizability if $v$ and $n$ are many-electron wave functions for the valence electrons only. 
Then the contribution from core electrons should be calculated separately. For the closed-shell core (or closed-shell atom or ion like Cu~II or Yb~III in the ground state) eq.~(\ref{e:pol}) can be reduced to
\begin{equation}\label{e:polc}
	\alpha_v(0)=\frac{2}{3}\sum_{c}\langle v|\hat d|\delta \psi_c\rangle,
\end{equation}
where $\hat d$ is the operator of the electric dipole moment and $\delta \psi_c$ is the RPA correction to the core state $c$ (see Eq.~(\ref{e:RPA})). The summation goes over all states in the core.

To calculate the polarizabilities of the clock states, we use the approach developed in Ref.~\cite{symmetry} for atoms or ions with open shells. It is based on Eq.~(\ref{e:pol}) and the Dalgarno-Lewis method~\cite{Dalgarno:1955}, which reduces the summation over the complete set of states to solving a matrix equation (see Ref.~\cite{symmetry} for details). This approach treats the $3d$ electrons in Cu~II and $4f$ electrons in Yb~III as valence electrons.
To calculate the contributions of the core electrons below the $3d$ or $4f$ shells, we use the Eq.~(\ref{e:polc}), in which summation over core state is limited to states below $3d$ or $4f$. To minimize the error in the difference between the ground state and clock state polarizabilities, we use the same approach for both states of both ions.

The results are presented in Table~\ref{t:pol}. Our results for the ground state polarizabilities are in excellent agreement with previous calculations.
To the best of our knowledge, the polarizabilities of the excited states of Cu~II and Yb~III ions were never calculated before.

	
The shift of the frequency of the clock transition due to black-body radiation (BBR) is given by~\cite{BBR}
	\begin{equation}\label{e:BBR}
	\delta \nu_{\rm BBR} = -1.6065\cdot10^{-6} \times T^4 \times\Delta \alpha(0),
	\end{equation}
where $T$ is a temperature (e.g., room temperature $T$= 300 $K$), $ \Delta \alpha(0)= \alpha_0({\rm CS}) - \alpha_0({\rm GS})$, is the difference between the clock state and ground state polarizabilities.
The calculated frequency shifts are presented in Table~\ref{t:pol}. The fractional BBR shifts for our Cu~II are close in value to those of
 Zn: $-2.5\cdot$10$^{-16}$, Cd: $-2.8\cdot$10$^{-16}$~\cite{Zn}, and Cu: $-3.4\cdot$10$^{-16}$~\cite{CuAgAu} and smaller than some other atomic clocks, such as Ca~\cite{Ca} and Sr~\cite{Sr} where fractional BBR shift is at the level of $10^{-15}$. As for the BBR shift in the Yb~III clock transition, its fractional value, $-5.95\cdot$10$^{-17}$, is one of the smallest among optical clock transitions. 

	\begin{table*}
	\caption{\label{t:pol}
		Scalar static polarizabilities of the ground states, $\alpha_0({\rm GS})$, and clock states, $\alpha_0({\rm CS})$,  and BBR frequency shifts for the clock transition of $^{63}$Cu~II and $^{171}$Yb~III.  $\delta\nu_{BBR}$/$\omega$ is the fractional contribution of the BBR shift; where $\omega$ is the clock transition frequency. 
	}

	\begin{ruledtabular}
		\begin{tabular}{cccccccc}
			&
			\multicolumn{2}{c}{$\alpha_0({\rm GS})$[$a_B^3$]}&
			\multicolumn{1}{c}{$\alpha_0({\rm CS})$[$a_B^3$]}&&
			\multicolumn{3}{c}{BBR, (\textit{T}= 300 K)} \\
			
			\cline{2-3}
			\cline{4-4}
			\cline{6-8}
			\multicolumn{1}{c}{Transition}&

			\multicolumn{1}{c}{Present} &
			\multicolumn{1}{c}{Other Cal.} &
			
			\multicolumn{1}{c}{Present} &

			\multicolumn{1}{c}{$\Delta\alpha$(0)}&
			\multicolumn{1}{c}{$\delta\nu_{BBR}$[Hz]}&
			\multicolumn{1}{c}{$\omega$[Hz]}&
			\multicolumn{1}{c}{$\delta\nu_{BBR}$/$\omega$} \\
			\hline
					\multicolumn{8}{c}{\textbf{Cu~II}}\\
				
2$\leftrightarrow$1 &5.36  &5.36 \footnotemark[1] &24.12&18.76&-0.1616&6.57$\cdot$10$^{+14}$ &-2.46$\cdot$10$^{-16}$ \\
				
3$\leftrightarrow$1 &5.36   &5.36 \footnotemark[1] &24.05&18.69&-0.1610&6.85$\cdot$10$^{+14}$&-2.35$\cdot$10$^{-16}$ \\
				
					\\
				\multicolumn{8}{c}{\textbf{Yb~III}}\\
				2$\leftrightarrow$1 &6.39   & 6.55 \footnotemark[2] &13.29&6.90&-0.0595&1.00$\cdot$10$^{+15}$&-5.95$\cdot$10$^{-17}$\\
				
			\end{tabular}
			\footnotetext[1]{~Ref.\cite{Johnson}}
			\footnotetext[2]{~Ref.\cite{pol_Yb2+}}

		\end{ruledtabular}
		
	\end{table*}

\subsection{ Zeeman Shift and Electric Quadrupole Shift}

Clock transition frequencies might be affected by external magnetic and electric fields. 
Zeeman shift caused by magnetic field strongly depends on whether the atom or ion has a hyperfine structure.
Both stable isotopes of copper ($^{63}$Cu and $^{65}$Cu) have non-zero nuclear spin ($I=3/2$) and non-zero hfs.
On the other hand, five stable isotopes of Yb have zero nuclear spins, and in two isotopes, spin is not zero (for $^{171}$Yb $I=1/2$, for $^{173}$Yb $I=5/2$). 
For atoms with zero nuclear spin, the first-order Zeeman shift can be avoided by considering transitions between states with $J_z=0$, while the second-order Zeeman shift is small due to the absence of the hfs.

Below we consider isotopes with non-zero nuclear spin,  $^{63}$Cu and  $^{171}$Yb.

The linear Zeeman shift is given by
\begin{equation}\label{e:z1}
\Delta E_{F,F_z} = g_F \mu_B B F_z,
\end{equation}
where $g_F$ is the $g$-factor of a particular hfs state. It is related to the electron $g_J$-factor by
\begin{equation}\label{e:f}
g_F=g_J \langle F,F_z=F,I,J|\hat J_z|F,F_z=F,I,J \rangle/F.
\end{equation}
Electron $g_J$ factors have values $g_3=1.32$, $g_2=1.16$ for Cu~II~\cite{NIST}, and $g_2=1.46$ for Yb~III (calculated value).
Linear Zeeman shift can be suppressed by averaging over the transition frequencies with positive and negative $F_z$.
 

Second-order Zeeman shift for transition between definite hfs components is strongly dominated by transitions within the same hfs multiplet.
Note that in this approximation, the shift is zero for the ground state (because $J=0$). For the clock states, the shift is given by 
\begin{eqnarray}\label{e:z2}
\delta E_{F,F_z}= \sum_{F'=F \pm 1,F'_z}\frac{|\langle F'F'_zIJ|\hat J_z|FF_zIJ\rangle x |^2}{\Delta E_{\rm{hfs}}(F,F')},
\end{eqnarray} 
where $x= g_J \mu_B B_m$ ( in which $g_J$ is electron $g$ factors, $\mu_B$  is the electron magnetic moment, and $B_m$ is a magnetic field), and  $\Delta E_{\rm{hfs}}(F,F')=E(F\,I\,J)-E(F'\,I\,J)$ is the hfs interval. For more detail, see Ref.~\cite{CuAgAu}. 

To calculate this shift, we need to know the hfs of the clock states. We calculated the hfs using the CIPT and RPA methods as described above.
The results for magnetic dipole hfs constants $A$ and electric quadrupole hfs constants $B$ are presented in Table~\ref{t:hfs}.
Using these numbers and Eq.~(\ref{e:z2}) we calculate the second-order Zeeman shift for all hfs components of the clock states of the $^{63}$Cu~II and  $^{171}$Yb~III ions. The results are presented in Table~\ref{t:Z2}. The shift is small and only slightly larger than in clock transitions of Cu, Ag, and Au~\cite{CuAgAu}. As in the case considered in Ref.~\cite{CuAgAu}, the shift can be further suppressed by taking appropriate combinations of the transition frequencies. It might be even easier here since we need to worry only about suppressing the Zeeman shift for the clock state while it is already strongly suppressed for the ground state.


	\begin{table}[!]
		\caption{\label{t:hfs}
		Hyperfine structure constants $A$ and $B$ in (MHz) of $^{63}$Cu~II and $^{171}$Yb~III ions. Nuclear spin \textit{I} of ($^{63}$Cu)= 3/2 and  \textit{I} of ($^{171}$Yb) = 1/2 , nuclear magnetic moment $\mu(^{63}{\rm Cu})=2.2236(4)\mu_N$ and $\mu(^{171}{\rm Yb})=0.49367(1)\mu_N$~\cite{Stone1};
			nuclear electric quadrupole moment $Q(^{63}{\rm Cu})=-0.220(15)~b$~\cite{Stone2} and $Q(^{171}{\rm Yb})=0$.} 
		
		\begin{ruledtabular}
			\begin{tabular}{cccccc}
				
				\multicolumn{1}{c}{No.}& 
				\multicolumn{1}{c}{Conf.}&
				\multicolumn{1}{c}{Term}&
				\multicolumn{1}{c}{\textit{E} (cm$^{-1}$)}&
				\multicolumn{1}{c}{hfs '\textit{A}' }&
				\multicolumn{1}{c}{hfs '\textit{B}' }\\
				\hline
				
				\multicolumn{6}{c}{\textbf{$^{63}$Cu~II}}\\
				1 & $3d^{9}4s$& $^3$D$_3$ &21932 &-186.46&-1.970\\
				2 & $3d^{9}4s$& $^3$D$_2$ &22733 &-34.62&-1.097\\
				\\
				\multicolumn{6}{c}{\textbf{$^{171}$Yb~III}}\\

				1 & $4f^{13}5d$&$^3${P}$\rm ^o_{2}$ &33350 &-41.46&0\\
			\end{tabular}
			
		\end{ruledtabular}
	\end{table}
	
\begin{table}
\caption{\label{t:Z2}
			Second-order Zeeman shifts $E_c$ (mHz/$(\mu \hbox{\rm T})^2$) for the clock states of $^{171}$Yb~III and $^{63}$Cu~II.}
		
\begin{ruledtabular}
\begin{tabular}{cccccc}
&&&
\multicolumn{3}{c}{($\Delta E_c$)/\textit{B}$^2_m$}\\
				
				\cline{4-6}
				
				&&&

				\multicolumn{2}{c}{$^{63}$Cu~II}&
				\multicolumn{1}{c}{$^{171}$Yb~III}\\
				
				\cline{4-5}
				\cline{6-6}

				\multicolumn{1}{c}{No.}&
				\multicolumn{1}{c}{$ F_c$}&
				\multicolumn{1}{c}{$F_{cz}$}&
				\multicolumn{1}{c}{$^3${D}$_{3}$}&	
				\multicolumn{1}{c}{$^3${D}$_{2}$}&
				
				\multicolumn{1}{c}{ $^3${P}$\rm ^o_{2}$}\\
				
				\hline
				1&1/2&$\pm$1/2&$-$& 9.127&$-$\\
				
				\\
				2&3/2&$\pm$1/2&0.8687& -5.967&1.021\\
				
				3&3/2&$\pm$3/2&0.5792& 2.107&0.6807\\
				\\
				4&5/2&$\pm$1/2&-0.3555& -2.265&-1.021\\
				5&5/2&$\pm$3/2&-0.1515& -1.360&-0.6807\\
				6&5/2&$\pm$5/2&0.2566& 0.4478&0.000\\
				\\
				7&7/2&$\pm$1/2&-0.3087& -0.8957&$-$\\
				8&7/2&$\pm$3/2&-0.2436& -0.7464&$-$\\
				9&7/2&$\pm$5/2&-0.1134&-0.4478&$-$\\
				10&7/2&$\pm$7/2&0.0818& 0.000&$-$\\
				\\
				11&9/2&$\pm$1/2&-0.2045&$-$&$-$\\
				12&9/2&$\pm$3/2&-0.1841&$-$&$-$\\
				13&9/2&$\pm$5/2&-0.1432&$-$&$-$\\
				14&9/2&$\pm$7/2&-0.0818&$-$&$-$\\
				15&9/2&$\pm$9/2&0.000&$-$&$-$\\
			\end{tabular}
		\end{ruledtabular}
	\end{table}
The electric quadrupole shift is due to the interaction of the atomic quadrupole moment Q with trapping electric field gradient, and a corresponding term in the Hamiltonian is
\begin{equation}
	H_{Q}=-\frac{1}{2} \hat Q_0 \frac{\partial \mathcal{E}_{z}}{\partial z}.
\end{equation}
Here $z$ is the quantization axis determined by the externally applied $B$ field. 
The spherical components of the quadrupole moment operator $\hat Q_m=|e|r^2 C^{(2)}_{m}$ are  the same as for the electric quadrupole (E2) transition.
The energy shift of a state with total angular momentum $J$ is proportional to the atomic quadrupole moment
of this state. It is defined as twice the expectation value of the  spherical component $Q_0=Q_{zz}/2$ of the quadrupole operator in the stretched state
\begin{equation}\label{eq:Q}
Q_J = 2\langle J,J_z=J|\hat Q_0| J,J_z=J \rangle.
\end{equation}
We calculate the values of $Q_J$ using the CIPT and RPA methods. The results are
$Q_{3}$ = 0.537 for the $^3$D$_3$ clock state of Cu~II, $Q_{2}$ = 0.299 a.u. for the $^3$D$_2$ clock state of Cu~II, $Q_{2}$ = -2.369 a.u. for the clock state of Yb~III. Note that $Q=0$ for the ground states of both ions because of the zero value of the total angular momentum $J$.

\subsection{Sensitivity of the Clock Transitions to Variation of the fine structure constant.}

\begin{table}[!]
\caption{\label{t:q}
Sensitivity of clock transitions to variation of the fine-structure constant ($q$, cm$^{-1}$ and $K=2q/E$) for clock transitions in Cu~II and Yb~III.} 
\begin{ruledtabular}
\begin{tabular}{cccccc}
				
\multicolumn{1}{c}{No.}& 
\multicolumn{1}{c}{Conf.}&
\multicolumn{1}{c}{Term}&
\multicolumn{1}{c}{$E_{\rm exp.}$}&
\multicolumn{1}{c}{$q$}&
\multicolumn{1}{c}{$K$}\\
\hline
\multicolumn{6}{c}{\textbf{Cu~II}}\\
1 & $3d^{9}4s$& $^3$D$_3$ &21929 &-4350 & -0.40\\
2 & $3d^{9}4s$& $^3$D$_2$ &22847 &-3700 & -0.32 \\
				\\
\multicolumn{6}{c}{\textbf{Yb~III}}\\
				
1 & $4f^{13}5d$&$^3${P}$\rm ^o_{2}$ &33386 &-42750 & -2.56\\
\end{tabular}
\end{ruledtabular}
\end{table}
	
Dependence of frequencies of atomic transitions on the fine structure constant in the vicinity of their physical values can be presented as	
\begin{equation}\label{e:q}
	\omega = \omega_0 + q\left[\left(\frac{\alpha}{\alpha_0}\right)^2-1\right]
\end{equation}
where $\alpha_0$ and $\omega_0$ are the present-day values of the fine structure constant and the frequency of the transition and $q$ are sensitivity coefficients that come from the calculations \cite{CJP}. When one atomic frequency is measured against another over a long period of time, their relative time-change is related to the time-change of $\alpha$ by
\begin{equation}\label{e:adot}
\frac{\dot \omega_1}{\omega_1} - \frac{\dot \omega_2}{\omega_2} = \left(K_1 - K_2 \right)\frac{\dot \alpha}{\alpha}.
\end{equation}

The dimensionless value $K=2q/\omega$ is usually called the {\em enhancement factor}.
To calculate $q$ (and $K$), we run computer codes at two different values of $\alpha$ and calculate the numerical derivative
\begin{equation}
q=\frac{\omega(\delta)-\omega(-\delta)}{2\delta},
\end{equation}
where $\delta =(\alpha/\alpha_0)^2-1$ (see Eq.~\ref{e:q}). The value of $\delta$ must be small to ensure linear behavior but sufficiently large to suppress numerical noise. Using $\delta =0.01$ usually gives accurate results. The calculated values of $q$ and $K$ for clock transitions of Cu~II and Yb~III are presented in Table~\ref{t:q}. As one can see, the sensitivity of the clock transitions of Cu~II to variation of $\alpha$ is not very high, 
so they may be used as anchor lines for a comparison with a high $K$ transition (see eq. (\ref{e:adot})). 
The sensitivity of the Yb~III clock transition is one of the highest among the systems considered so far. It is close to the sensitivities of recently suggested clock transitions in Yb~\cite{DFS-Yb} and Au~\cite{CuAgAu} and slightly smaller than the sensitivity of the most sensitive clock transitions in Yb~II and Hg~II~\cite{CJP}.

	
\section{Conclusion}
We have investigated a possibility to use Cu~II and Yb~III ions  as optical ion clocks of high accuracy. Energy levels, lifetimes, transition rates, scalar static polarizabilities of the ground and clock states, and the  BBR shifts have been calculated using the CIPT method. We have obtained a good agreement with  previous data that are available to compare. Sensitivity to "new physics" such as variation of the fundamental constants has been studied. 
 The uncertainty estimates  for the Yb~III ion and its high sensitivity to new physics indicate that  Yb~III atomic clock may  successfully compete with the latest generation of clocks.
 
 \section{ACKNOWLEDGEMENTS}
 
This work was supported by the Australian Research Council Grants No.  DP190100974 and  DP200100150.

	\end{document}